\documentclass[preprint]{aastex}

\newcommand \gta {\mathrel{\vcenter
     {\hbox{$>$}\nointerlineskip\hbox{$\sim$}}}}

\newcommand\kms{km~s$^{-1}$}
\newcommand\gr{$\gamma$-ray}

\newcommand\grb{$\gamma$-ray burst}
\newcommand\grbs{$\gamma$-ray bursts}

\begin{document}

\title{GRB~021004: a Massive Progenitor Star Surrounded by Shells}

\author{Bradley E. Schaefer, C. L. Gerardy, P. H\"oflich, A. Panaitescu,
R. Quimby, J. Mader, G. J. Hill, P. Kumar, J. C. Wheeler} \affil{Astronomy
Department, University of Texas, Austin, TX 78712}
\email{schaefer@astro.as.utexas.edu, gerardy@astro.as.utexas.edu,
pah@astro.as.utexas.edu, adp@astro.as.utexas.edu,
quimby@astro.as.utexas.edu, jmader@astro.as.utexas.edu,
hill@astro.as.utexas.edu, pk@astro.as.utexas.edu,
wheel@astro.as.utexas.edu}

\author{M. Eracleous, S. Sigurdsson, P.  M\'esz\'aros, B. Zhang}
\affil{Department of Astronomy, Pennsylvania State University}
\email{mce@astro.psu.edu, steinn@astro.psu.edu, nnp@astro.psu.edu, 
zhang@astro.psu.edu}

\author{L. Wang}
\affil{Lawrence Berkeley Laboratory}
\email{lifan@panisse.lbl.gov}

\author{F. V. Hessman}
\affil{Universit\"ats-Sternwarte, 
Georg-August-Universit\"at G\"ottingen}
\email{hessman@uni-sw.gwdg.de}

\and

\author{V. Petrosian}
\affil{Stanford University}
\email{vahe@astronomy.stanford.edu}

\begin{abstract}

We present spectra of the optical transient of GRB~021004 obtained
with the Hobby-Eberly Telescope starting 15.48, 20.31 hours, and
4.84 days after the \grb\ and a spectrum obtained with the H. J.
Smith 2.7 m Telescope starting 14.31 hours after the \grb.  
GRB~021004 is the first burst afterglow for which the spectrum is
dominated by absorption lines from high ionization species with
multiple velocity components separated by up to 3000 \kms.  We argue
that these absorption lines are likely to come from shells around a
massive progenitor star.  The high velocities and high ionizations
arise from a combination of acceleration and flash-ionization by the
burst photons and the wind velocity and steady ionization by the
progenitor.  We also analyze the broad-band spectrum and the light
curve so as to distinguish the structure of gas within 0.3 pc of the
burster.  We delineate six components in the medium surrounding the
\grb\ along the line of sight:  (1) The $z \cong 2.293$ absorption
lines arise from the innermost region closest to the burst, where
the ionization will be highest and the 3000 \kms\ velocity comes
from the intrinsic velocity of a massive star wind boosted by
acceleration from the burst flux.  For a mass loss rate of $\sim 6
\times 10^{-5} M_\odot {\rm yr}^{-1}$, this component also provides
the external medium with which the jet collides over radial
distances 0.004--0.3 pc to create the afterglow light.  (2)  A
second cloud or shell produces absorption lines with a relative
velocity of 560 \kms. This component could be associated with the
shell created by the fast massive star wind blowing a bubble in the
preceding slow wind at a radial distance of order 10 pc or by a
clump at $\sim$ 0.5 pc accelerated by the burst.  (3) More distant
clouds within the host galaxy that lie between 30-2500 pc and have
been ionized by the burst will create the $z \cong 2.33$ absorption
lines.  (4-6)  If the three bumps in the afterglow light curve at
0.14, 1.1, and 4.0 days are caused by clumps or shells in the
massive star wind along the line of sight, then the radii and
over-densities of these are 0.022, 0.063, and 0.12 parsecs and 50\%,
10\%, and 10\% respectively.  The immediate progenitor of the \grb\
could either be a WC-type Wolf-Rayet star with a high velocity wind
or a highly evolved massive star the original mass of which was too
small for it to become a WN-type Wolf-Rayet star.  In summary, the
highly ionized lines with high relative velocities most likely come
from shells or clumps of material close to the progenitor and these
shells were plausibly produced by a massive star soon before its
collapse.

\end{abstract}


\section{Introduction}

The large absorption line red shift measurements settled the issue that
most \grbs\ are at cosmological distances.  The physical mechanism of the
origin of the \grbs, questions of whether there are more than one
mechanism, and the issue of whether \grbs\ interact only with the
interstellar medium or with a circumstellar environment are still very
open. Correlations with star formation and observations of putative iron
lines with the Chandra observatory \citep{piro00} have put new focus on
the connection to a supernova-like core collapse event.

Currently, most observations of \grb\ afterglows are photometric,
with the effort being to find the power law behavior of the
afterglow brightness as a function of wavelength and time.  Most of
the 25 known red shifts have been measured by deep spectroscopy on
the host galaxy long after the afterglow has faded.  Prior to
GRB~021004, only eight \grb\ afterglows have had early-time spectra
that reveal absorption lines from the host galaxy (e.g., Vreeswijk
et al. 2001).  Of these, virtually all have been low excitation
lines in one velocity system that are likely to arise in gas far
from the burster.  Two exceptions to this are GRB~011211 which shows
\ion{C}{4} and \ion{Si}{4} in addition to lower excitation lines
\citep{holland02}, and GRB~010222 which displayed two low-ionization
systems separated by $\sim 119$ \kms\ that could arise from clouds
at different positions within the host galaxy \citep{mirabal02a}.  
GRB~020813 has a spectrum which displays \ion{C}{4} absorption lines
in both a z=1.223 and z=1.255 systems \citep{barth02}.  With so
little known, and many afterglows revealing new properties, more
prompt spectroscopy is clearly needed.

Afterglow spectra could show absorption lines arising from the host
galaxy (especially the Fe and Mg ISM lines) that will give the red shift.  
The red shift is the key parameter for determining the burst energetics
and hence further testing the conclusions of \citet{frail01} and
\citet{pk01} that \grbs\ have a nearly constant total energy.  One can
also improve the calibration (currently based on 9 events, see also
Schaefer, Deng, \& Band 2000) that the spectral lags and light curve
variability are Cepheid-like luminosity indicators that could turn \grbs\
into a premier tool for cosmology out to z$\sim$10 \citep{schaefer02}.  
Intrinsic afterglow spectral features of any kind would be a boon to
interpretation.  If there are underlying supernova-like explosions,
corresponding spectral features could be superposed on the continuum and
revealed by careful subtraction of the well-defined power-law continuum.  
Detection of, for instance, features of red shifted Ca H \& K, which tend
to be among the strongest lines in all types of supernovae, would yield
both a red shift and clues to the nature of the explosion itself.  Other
possibilities would be to discover Fe emission lines, which would yield
temperatures, red shifts, and abundances.  Perhaps the most important
possibility is that of serendipitous discovery of some unexpected
feature.

GRB~021004 provided an unprecedented opportunity to obtain prompt
spectroscopy of the early afterglow of a \grb.  GRB~021004 was discovered
by instruments on the HETE II satellite at 12:06 UT on 2002 October 4
\citep{shirasaki02}.  Prompt discovery of the optical transient by
\citet{fox02a} allowed rapid spectroscopic observations.  \citet{fox02b}
obtained the first spectra and identified two intervening systems at z =
1.38 and 1.60 from \ion{Mg}{1} and \ion{Mg}{2} absorption.  
\citet{erac02} confirmed those features, noted several \ion{Fe}{2}
absorptions at these redshifts, and pointed out four absorption lines at
$\sim$ 4633, 4664, 5109, and 5152 \AA\ (see also Sahu et al. 2002;
Castander et al. 2002).  These were identified by \citet{chorfil02} as
\ion{C}{4} and \ion{Si}{4} features at red shift of $\sim 2.3$.  
Chornock \& Filippenko also identified Ly$\alpha$ emission at z = 2.323
and absorption components and perhaps Ly$\beta$ at similar red shifts.  
\citet{salamanca02} identified four absorption components of \ion{C}{4}
at z = 2.295, 2.298, 2.230 and 2.237, noting that the total spread is
about 3000 \kms.  They remarked that this velocity dispersion would be
difficult to interpret as due to a single cluster of galaxies and noted
that if associated with a supernova shell the supernova must have
substantially preceeded the \grb\ to be optically thin.  
\citet{mirabal02b} suggested the identification of one absorption feature
as low excition \ion{Al}{2} at z = 2.328 and identified that, and the
Ly$\alpha$ emission, as the red shift of the host galaxy.  They raised
the issue of the velocity dispersion being due to a wind in a Wolf-Rayet
like progenitor. \citet{djorg02} noted that the Ly$\alpha$ flux may imply
a strong star formation rate in the host galaxy.  \citet{savaglio02}
reported a very high resolution spectrum (R = 50,000) that revealed the
complex fine structure of many of the high ionization lines.  They
determined that the individual \ion{C}{4} features show complex flow
spanning about 1000 \kms ~ and conclude that the $z \cong 2.326$ system
is probably close to the \grb. They note that of the four high red shift
systems, the low ionization lines, \ion{Al}{2} and possibly \ion{Si}{2},
only occur in the highest red shift, $z \cong 2.326$, component.  
\citet{matheson02} and M{\o}ller et al. (2002) have provided an excellent
compendium of line identifications along with detailed line profiles.  
\citet{matheson02} give a spectral index in the optical of $F_{\nu}
\propto \nu^{0.96 \pm 0.03}$, and report a reddening of the continuum in
the blue region of the spectrum over the first three days.

Here we report spectra obtained within the first day after outburst and
five days later.  We concentrate on constraining the circumburst medium
in the context of the high ionization lines and the afterglow light
curve.

\section{Observations}   

Spectra of the optical afterglow were obtained with both the 2.7 m Harlan
J. Smith Telescope using the IGI imaging spectrograph \citep{hill03} and
the 9.2 m Hobby-Eberly Telescope (HET) with the Marcario Low-Resolution
Spectrograph \citep{hill98} at McDonald Observatory. Three consecutive
exposures of 900 seconds each were taken with the 2.7m telescope beginning
14.31 hours after onset of the \grb, followed by sets of three 900 second
exposures taken with the HET beginning at 15.48 hours, 20.31 hours, and
4.84 days after the burst.  A 2.5 arcsecond slit was used for the 2.7m
data, while the HET observations utilized a 2.0 arcsecond slit, giving all
data sets a 16 \AA\ resolution. The spectra were reduced using
IRAF\footnote{IRAF is distributed by the National Optical Astronomy
Observatories, which are operated by the Association for Research in
Astronomy, Inc., under cooperative agreement with the National Science
Foundation}, and wavelength calibrated using Argon, Neon, Cadmium, and
Mercury lamps.  The combined spectrum from the first night is displayed in
Figure 1.

Our first night spectrum shows a collection of absorption lines longward
of 5800 \AA\ that arise from iron and magnesium of relatively low
ionization at red shifts of 1.38 and 1.60.  These are due to intervening
galaxies with no relevance to the \grb\ analysis.  The lines at red
shifts of $z \sim 2.3$ are presented with their measured properties in
Table 1.  The lines are separated into two primary velocity components,
with $z \cong 2.326 \pm 0.001$ and $z \cong 2.293 \pm 0.003$.  We
identify the 4587.2 \AA\ line as being from \ion{Si}{4} (c.f. Matheson et
al. 2002).  There are few resonance lines from the ground state of
highly-ionized elements of high cosmic abundance in the observed
wavelength range, and the only one that has not produced a detected
absorption line is \ion{N}{5} with rest wavelengths 1238.8 and 1242.8
\AA\ for which we can only present an upper limit to its equivalent
width.

The separation of the two high redshift primary velocity components
corresponds to a relative velocity of 3000 \kms\ in the frame of the
host galaxy.  The unblended lines for the $z \cong 2.326$ velocity
component are significantly resolved in our spectrum, with a FWHM of
$\sim 1000$ \kms.  From our own data, this implies either multiple
components or one component with a velocity range that large. Other
spectra \citep{savaglio02,mirabal02b,salamanca02,moller02} have
resolved the high red shift component into at least two
sub-components with a velocity separation of 560 \kms.  Thus, the $z
\sim 2.3$ system consists of at least three separate velocity
components, with velocities of 0 \kms, 560 \kms, and 3000 \kms\
towards Earth relative to the highest red shift component.

\citet{draine02} presented detailed calculations of the complex of
absorption lines produced from 1110-1705 \AA\ by vibrationally excited
molecular hydrogen in a molecular cloud near a \grb.  Individual lines
have a typical depth of 20\% for normal low-resolution sprecta and might
be difficult to detect without very high signal.  Fortunately, the entire
pattern of lines is characteristic of the process and so all of the lines
(with their unique positions and depths) can be simultaneously sought in
any spectrum.  The idea is that a spectrum in which any one line would be
highly insignificant might yet have the presence of many lines detectable
at a high confidence level.  Even a noisy spectrum might have its
apparent noise following the characteristic molecular hydrogen features
hence resulting in a significant detection.  Many of the individual lines
are extremely saturated with optical depths of $\>300$ in the line center
and with equivalent widths of 0.1 \AA\ or greater, so that the strength
of the absorption is quite insensitive to the column density of
vibrationally excited molecular hydrogen, $N(H_2^*)$.  B. Draine has
kindly provided us with the transmission spectrum for excited molecular
hydrogen for our spectral resolution.  We have used this as a template to
cross correlate with our spectrum.  For a red shift of 2.3, we cover the
range from 1240-1705 \AA.  We find no significant correlation peak for
any red shift of the template.  Within the red shift range of 2.29--2.33,
the largest peak in the cross correlation is never more than the RMS of
the cross correlation over a much wider red shift range.  The implication
of this negative result will be discussed in Section 3.5.

We found no significant differences among our spectra within the first
night.  Specifically, the equivalent widths of the absorption lines were
all identical to within the errors.  We have constructed broad-band
photometry from our reduced spectra using the photometric pass-bands of
Landolt (1992) and normalizing the colors to match the observations
concurrent to the 2.7m spectrum.  The resulting synthetic photometry for
the B, V, and R magnitudes yields colors that are constant to within 0.04
mag for the first night.

For a comparison between our observations on the first and fifth nights,
we are handicapped by the poor signal on the fifth night due to the
faintness of the afterglow.  The only lines for which we can quote line
widths on the fifth night are the two \ion{C}{4} lines which have
equivalent widths in the rest frame of the host (the observed width
divided by 1+z) of $1.0 \pm 1.6$ \AA\ and $5.7 \pm 2.4$ \AA\ at the same
wavelengths as for the first night.  This is consistent with the first
night.  We confirm the report by \citet{matheson02} that the blue end of
the continuum became redder from the first night to subsequent nights.  
In particular, our synthetic photometry shows the B-V color to have
reddened by $0.25 \pm 0.13$ mag from the first to the fifth night, while
the V-R color was unchanged, $0.01 \pm 0.07$ mag, over the same interval.

\section{The High Ionization Features}

\subsection{The Source of the Absorbing Gas}

The Lyman alpha {\it emission} line is likely to represent the average
velocity of the host galaxy, although its exact wavelength will shift
redward somewhat due to the absorption component.  The red shift of the
Ly$\alpha$ line is variously given as 2.3351, 2.332, 2.328, and 2.323
\citep{moller02,matheson02,mirabal02b,chorfil02}.  The average velocity of
the host galaxy is also likely represented by the low-ionization feature
of \ion{Al}{2} at 2.328.  The highest red shift components of the
\ion{C}{4} and \ion{Si}{4} lines correspond to this host galaxy velocity.  
The velocity of the burster itself might vary by up to several 100 \kms\
from the average velocity for the host.  The other primary components of
these high-ionization lines are at velocities of 560 \kms\ and 3000 \kms\
blueward of the host galaxy velocity.

The presence of the $z \sim 2.3$ lines of \ion{C}{4} and \ion{Si}{4}
shows that the absorbing gas is highly ionized.  Such lines are commonly
seen in the spectra of distant quasars through intervening galaxies where
the ionized material is in a collisionally ionized halo.  Could the
absorbing gas for GRB~021004 just be in halos of chance galaxies along
the line of sight? The presence of three velocity components within the
$z \sim 2.3$ system with separation of up to 3000 \kms\ argues strongly
that this cannot be.  The probability of two galaxies lying in so small a
red shift range just blueward of the host is rather small even in a
cluster of galaxies, and the velocities are too large to allow for a
bound cluster.  Sargent, Steidel, \& Boksenberg (1988) present statistics
on lines with rest frame equivalent widths of more than 0.3 \AA\ (i.e., a
threshold much smaller than the observed lines for GRB~021004), and find
that absorption systems with $z \sim 2.3$ occur with a rate of $\sim 1.5$
per red shift unit.  The probability of getting two absorbers within a
range of 0.033 in $z$ in front of the host galaxy is thus 0.0024.  Such
an occurrence is not impossible (c.f. the $z \sim 2.85$ system for
Q1511+091, Sargent, Steidel, \& Boksenberg 1988), but it is too small to
be plausible for GRB~021004.  It is also improbable that random
intervening galaxies will have equivalent widths as thick as 1.4 \AA\ (as
for the observed \ion{C}{4} lines from the systems blueward of the host
galaxy).  The distribution of equivalent widths (for both members of the
\ion{C}{4} doublet) falls off exponentially such that the probability for
any one galaxy being above this threshold is $\sim 0.2$, or 4\% for both
systems.  In all, the probability of getting two chance galaxies with a
red shift within 0.033 of the host and with rest frame equivalent widths
of $\ge 1.4$ \AA\ is 0.00010. We conclude that the high-ionization
absorption lines are not from intervening galaxies.

The only other situation for which the \ion{C}{4} and \ion{Si}{4} lines
are seen in absorption is when gas is subjected to a high flux of
ionizing radiation, as in an AGN, a Wolf-Rayet star, or a \grb.  An AGN
at the position of the \grb\ would already have been detected, and so
this possibility can be rejected.  With only $\sim 3100$ Wolf-Rayet stars
in our own Galaxy, the odds of the line of sight passing close to 1--3
such stars is small (ignoring clustering) provided that the star is not
the progenitor of the \grb.  That the progenitor is itself a Wolf-Rayet
star is plausible, because the progenitor is likely to be a massive star
just before a core-collapse.  Alternatively, the ionizing radiation could
come from the \grb\ itself.  Thus, we think that the most likely means to
ionize the gas to high levels is either radiation from the burst
progenitor or the burst itself, with both cases requiring the ionized gas
to be physically associated with the burster.

\subsection{Ionizing the Gas}

Wolf-Rayet stars have temperatures from 30,000--70,000 K and supergiant
luminosities.  For the hotter stars, this is sufficient to ionize the
surrounding nebulosity to levels where a significant fraction of the
carbon and silicon atoms can produce \ion{C}{4} and \ion{Si}{4}
absorption and emission.  Thus, a Wolf-Rayet-like progenitor might
already ionize the surrounding gas such that later light from the
afterglow will have silhouettes of these high-excitation lines.

Alternatively, the \grb\ itself is an obvious source of a huge ionizing
flux.  The most frequent mode of ionization of carbon and silicon will be
for the \gr\ to knock out an inner electron, whereupon autoionization
will cause ejection of the valence electrons.  In addition the ejected
electrons will have sufficient energy to ionize other atoms.  In the
remainder of this section, we will estimate a crude distance range over
which the burst flux can ionize enough gas so as to produce the observed
absorption lines.

The cross section for photoelectric interactions varies strongly with
photon energy, so the probability that any given atom is ionized by burst
flux requires an integral over energy.  The cross sections for carbon and
silicon are tabulated in \citet{hkm92} along with a prescription for
interpolation.  For carbon, the cross sections are 0.0213, 40.3, and
44500 barns (one barn equals $10^{-24} ~ {\rm cm}^2$) at photon energies
of 100, 10, and 1 keV respectively.  With this, we see that most of the
ionization will be from burst photons in the x-ray band.

The burst spectrum has only been measured down to a photon energy of 7
keV (23 keV in the frame of the host galaxy) by HETE-2.  Above 7 keV,
\citet{lamb02} state that the spectrum is well fit by a single power law
such that $dN/dE \propto E^{-1.64}$ and has a fluence from 7--400 keV of
$3.2 \times 10^{-6} ~ {\rm erg~cm}^{-2}$.  Below 7 keV, the spectrum will
likely turn over to asymptote at a slope close to the synchrotron limit
of $dN/dE \propto E^{-2/3}$.  Here, we will model this behavior as a
broken power law with indices of -1.64 and -0.66 above and below a break
energy, $E_{break}$.  We also impose a lower energy cutoff in our
calculations at $E_{cutoff}$, below which we assume that no flux reaches
the gas.  This is to model the possible photoelectric absorption of
photons before they reach the gas.  We normalize this spectrum to the
7--400 keV fluence reported by HETE-2 and correct the photon energies of
this observed spectrum by a factor of $1+z$ to get the spectrum as seen
by gas in the host galaxy frame.

The expectation value for the number of times that a given atom will be
ionized by burst flux is $\int \int (dN/dE) ~ \sigma ~ dt~ dE$, where
$dN/dE$ is the burst photon spectrum (in units of photons ${\rm cm}^{-2}
{\rm s}^{-1} {\rm keV}^{-1}$) as a function of distance from the burster,
$\sigma$ is the cross section for ionization as a function of photon
energy $E$, and $t$ is the time.  For these calculations, we adopt
$E_{break} = 7$ keV.  We find that the result is highly sensitive to the
value of $E_{cutoff}$ (due to the large cross section for ionization at
low energies), which itself is highly uncertain since we do not know the
degree of absorption of the low energy photons before the radiation hits
the gas.  For $E_{cutoff} = 0.3$ keV in our rest frame (1 keV in the host
galaxy), we expect $\sim 100$ ionizing interactions per carbon atom at a
distance of 1 pc.  For $E_{cutoff} = 0.03$ keV in our rest frame (0.1 keV
in the host galaxy), we expect $\sim 600$ ionizing interactions per
carbon atom at a distance of 1 pc.  The expected number of ionizing
interactions will scale as the inverse square of the distance.  For this
range of cases, the expected number of ionization interactions is unity
at distances ranging from 10-25 pc.  This provides an order-of-magnitude
estimate of the minimal range out to which the burst flux will completely
ionize carbon.

The radius at which significant high ionization absorption can be caused
may be substantially farther than this calculated distance for two
reasons.  First, only a small fraction of the carbon need be ionized to
result in significant absorption.  Since the cross section for the
\ion{C}{4} resonance line is large, the intervening gas need only have a
column density of $10^{14} {\rm cm}^{-2}$ in the \ion{C}{4} state to
account for the observed absorption line.  This corresponds to a fraction
of $\sim 0.0003$ of the total carbon expected in a smooth wind (see \S
3.4).  If only 0.01\% of the carbon must be ionized to \ion{C}{4} to
create the necessary column density, then the burst's effective radius
for ionization is larger than the previous distance estimate by a factor
of 100.  This factor depends on the column density of the cloudy ISM.  
Second, the electrons ejected from the neutral atoms at high velocity
will also have a high cross section for ionizing further atoms.  This
effect depends on the density of the gas and the spectrum of the
electrons \citep{lotz67}. This effect will roughly increase the ionizing
radius by a factor of the square root of the ratio of the electron energy
to the ionization energy, which might be up to a factor of order ten.  
In all, there are many uncertainties and dependencies on unknown
conditions, but the effective ionization radius of the burster might be
of order 10--100 times larger than calculated in the previous paragraph.  
This would imply that any gas cloud within perhaps 100--2500 pc of the
burster will suffer enough ionization to create a sufficient column
density of \ion{C}{4} so as to form a detectable absorption line.

The complementary question is how close to the burster some gas will be
incompletely ionized at the epoch of observation.  This is important
since we later consider the possibility of the absorbing gas being quite
close ($\sim 0.2$ pc) to the burst.  \citet{laz02} have presented a
detailed and extensive calculation that shows that even a very thick wind
will have all of its hydrogen completely ionized by a GRB flash on time
scales of milliseconds or faster.  The implication is that most (if not
all) of the atoms of carbon in a wind will also be completely ionized,
and thus there may be no atoms in the \ion{C}{4} state to create
absorption lines near the burst.

Whereas Lazzati \& Perna considered a smooth wind, the column density of
\ion{C}{4} $\sim 10^{14}$ cm$^{-2}$ could be accounted for by clumps in
the gas.  The existence of such clumps is plausible since images of
Wolf-Rayet winds show structure on all size scales.  Also, the
quasi-stationary flocculi (clumps of gas created in the Wolf-Rayet
progenitor wind) in Cas A and the Kepler supernova remnants have
densities of $1-2 \times 10^{4} {\rm ~cm}^{-3}$ \citep{ger01} after
correcting by a factor of 4 for the shock jump conditions, with a typical
projected size of $5 \times 10^{16}$ cm in width and $15 \times 10^{16}$
cm in length.  For GRB~021004 in particular, the existence of clumps is
indicated by the three bumps in the afterglow light curve (see \S 5)
where the clumps might be large in size with moderate contrast or might
be small in size yet with large central densities.  The existence of
clumps in the winds of massive stars is thus expected, and these clumps
can have densities $\gta$ $10^4 {\rm cm}^{-3}$.

Clumps can serve in two ways to allow for the existence of \ion{C}{4}
close to the \grb.  First, the clump will shield portions of the wind
material (both within the clump and within the shadow of the clump) from
the high burst flux.  For example, a clump with size $10^{17}$ cm with a
density of $10^4 {\rm~cm}^{-3}$ at a distance of a fraction of a parsec
from the burster should absorb the ionizing flux.  Second, the higher
density within clumps can allow for faster recombination that can
generate \ion{C}{4} a day after the burst even if the burst had
completely ionized the carbon.  The degree of recombination that will
have occured when we see the absorption a day or so after the burst has
ionized the gas depends strongly on the density.  For ionized carbon, the
recombination time scale is $\sim 10^{5}$ days for densities of $10
{\rm~cm}^{-3}$ that would be typical of a Wolf-Rayet wind at a distance
of $\sim 0.1$ pc.  The recombination time scale would be $\sim 100$ days
for densities of $10^4 {\rm~cm}^{-3}$ that are typical of the densities
observed in supernova remnant knots that are presumed to have formed in
the progenitor's wind.  Within such clumps, the fraction of carbon in the
\ion{C}{4} state should be $\sim 0.01$ a day after the burst.  A clump
with density $10^4 {\rm~cm}^{-3}$ and size $10^{17}$ cm will produce a
column of $10^{19} {\rm cm}^{-2}$ in \ion{C}{4}.  This is $10^5$ times
larger column than is required to produce the observed absorption lines.  
Detailed calculations of the ionization structure are difficult without
knowledge of the density structure in the wind.  Nevertheless, it appears
that shadows and recombination in the expected clumps can yield a
\ion{C}{4} column density of $10^{14} {\rm cm}^{-2}$ even quite close to
the burster.

	The calculation of the ionization structure is not straight
forward even if the density structure is given.  One trouble is that the
burst spectrum below 7 keV is unknown, and this is the critical energy
region for ionizing the gas due to the large cross sections for
photo-ionization.  A related problem is accounting for the ionizing flux
from the afterglow, which is comparable to that from the burst itself.  
An additional problem is that stimulated recombination will ensure that
some fraction of the carbon will not be completely ionized even in a
radiation field of the highest intensity.  Yet another issue is that the
photo-ionized electrons likely have roughly a power-law energy
distribution, whereas the cross sections for ordinary and stimulated
recombination are calculated for thermal electron distributions.  The
correct recombination rates are likely to be significantly different than
the ``usual"  calculations.  In principle, these problems can all be
surmounted by substantial theoretical effort.  In the mean time, we
conclude that it is difficult to make any definite statements regarding
the ionization state of gas close to the burster.

A cloud far from the \grb\ will have a low relative velocity
corresponding to ordinary galactic rotation.  Such a cloud would be an
obvious candidate for creating the high-excitation low-velocity
absorption component.  Another obvious candidate to contribute to the
high-excitation low-velocity lines is the shell piled up by the O main
sequence star wind as it blows a bubble in the local ISM at a distance of
perhaps 30 pc.  Thus, there are two adequate sources of material along
the line of sight with distances from 30--2500 pc that can be ionized by
the burst flux.

\subsection{Accelerating the Gas}

Gas near the burster will receive a tremendous blast of photons, and
interactions will transfer some of the photons' momentum to the gas.  The
total radiative acceleration of the gas that determines its final velocity
will occur during the few seconds when the burst flux passes through the
gas a few hours before the time at which the absorption lines are produced
by the gas.  How much momentum gets transferred depends on the cross
section of the gas, the spectrum of the burst, and the flux of the burst
photons.  The final velocity will fall off as the inverse-square of the
distance from the burst.  At some distance, the blast will accelerate the
gas to 3000 \kms, and this process could account for the observed high
relative velocity of one of the absorption components.  In this
subsection, we present a detailed calculation of the radiation-induced
velocity as a function of distance from the burster.

The radiative acceleration $g_r$ (as a function of distance R and time
t) by a central energy source is given by 
\begin{equation}
g_r(R,t) = 4 \pi c^{-1} \int \sigma(E) ~ m_n^{-1} ~ H_E(R,t) ~ dE,
\end{equation}
where $\sigma(E)$ is the
cross section for interaction with a burst photon of energy $E$, $m_n$
is the mass per nucleon, and $H_E$ is the Eddington flux which is the
first moment of the intensity (with units of erg cm$^{-2} ~ {\rm
s}^{-1} ~ {\rm erg}^{-1})$.  The velocity due to radiative acceleration
is 
\begin{equation}
v = v_0 + \int_{\Delta t} g_r dt 
\end{equation}
where $v_0$ is the initial
velocity of the gas, and $\Delta t$ is the time interval covering the
burst duration as the photons pass through the gas being accelerated.  
This integral over time will change the burst flux into a fluence,
which is to say that the acceleration only depends on the number of
incident photons and not on their time distribution.  This change in
velocity structure in the gas around the burst will not lead to any
significant changes in the density structure on time scales for which
we are concerned.  That is, a parcel of gas accelerated to, say, 3000 
\kms\ will move only 1.7 AU in one day, so as to cause a $\sim 0.03 \%$ 
density change within a gas clump of size, for example, $10^{17}$ cm  
if the backside suffers zero acceleration.

The \gr\ cross sections for the photon-matter interaction have been
included as detailed in H\"oflich (1991).  Because the photon energies far
exceed the binding energies of the electrons in atoms, the interaction of
electrons and photons is given by the Klein-Nishina cross section per
nucleon.  For lower energies, we determine the angle dependent
Klein-Nishina cross section by a standard rejection technique and obtain
the cross section by formal integration over angle.  Bound-free cross
sections have been included according to Veigel (1973).  The X-ray
opacities are dominated by bound-free transitions from inner shell
electrons (which are always populated since when an inner shell electron
is ejected it is rapidly filled by one of the outer electrons), and are
independent of the conditions in the gas cloud.  Thus, the bound-free
opacity is rather model independent.

The final velocities are a function of composition and the burst spectrum.  
For composition, we adopted either a normal solar abundance \citep{ag89},
a Wolf-Rayet composition \citep{wlw93}, or a pure gas of ionized hydrogen.  
For the spectrum, we adopted broken power laws with a low-energy cutoff,
as described in the previous subsection.  The results for $v_0 = 0$ are
displayed in Figure 2.  We find that the composition of the gas has little
effect.  We find that the flux around one keV (in the frame of the host
galaxy) is very important, due to the large cross sections associated with
bound-free interactions.  Unfortunately, the flux for energies below 7 keV
are not known and we can only assume plausible spectral shapes, as
discussed in the previous section.

Our favorite model of the final velocity as a function of the radial
distance from the burster is represented by the thick line in Figure
2.  To accelerate gas from rest to 3000 \kms\ implies that the gas
is at a distance of 0.2 parsec.  If the gas were already expanding
at a velocity of 2500 \kms, then the radial distance of the gas
would be $\sim$ 0.5 parsec. Gas initially at rest at $\sim$ 0.5 pc,
would be accelerated to $\sim$ 500 \kms, corresponding to one of the
observed absorption components.

If the cloud has a substantial extent along the line of sight, then the
inner edge will be accelerated to a higher velocity than the outer edge.  
This would result in a broad line, instead of the narrow lines observed.  
The radiative acceleration mechanism thus requires that the intervening
cloud be physically narrow.  This could arise from a small cloud (as a
clumpy part of a wind) or from a thin shell (presumably created by the
progenitor).  For quantitative limits, the width of the carbon lines
appears to be less than 400 \kms, for which the dominant absorbing region
in the shell must be thinner than 7\% of its radius provided the gas is
accelerated from rest to around 3000 \kms\ by the burst radiation.  If the
gas already had an initial velocity of 2000 \kms, then the dominant
absorbing region in the shell must be thinner than 21\% of its radius.  
If an absorbing gas has a wind-like density distribution (that falls off
as the inverse-square of the radial distance), then 50\% of the absorption
will come from between the inner edge of the absorbing region to twice
that radius.  Thus, for absorption from a smooth wind, the radiative
acceleration must be small compared to the initial wind velocity to avoid
excessively broad absorption lines.  However, if the absorbing gas has a
significant clump superposed on a wind-like structure (see \S 5),
then such a clump may be sufficiently thin to form a narrow absorption
line despite a large radiative acceleration even in a wind.

This acceleration of gas close to the burster must occur, the only
question is the size of the effect.  This will depend primarily on
unknown details of the low energy photons in the burst spectrum
after they have passed through material at small radial distances.  
For reasonable assumptions ($E_{cutoff} \ll 10$ keV), the radiative
acceleration will induce velocity changes of 1000 \kms\ or greater
for gas within a fraction of a parsec from the burster.  This
accelerated gas must include that which is creating the afterglow,
as well as the dense gas interior to the afterglow shock that
represents the inner region of any wind that was driven off the
progenitor.

\subsection{Massive Star Winds}

Wolf-Rayet stars are very hot (30,000-70,000 K) stars shedding a thick
wind for which the characteristic spectra involve bright and broad
emission lines with P Cygni profiles.  They are highly evolved stars, the
main sequence masses of which were larger than 35 $M_\odot$ or so.  For
the most massive stars ($\sim 60 M_\odot$) the evolutionary path is from
an O main sequence star to a Luminous Blue Variable to a Wolf-Rayet star
of the WN class to a Wolf-Rayet star of the WC class to a supernova
collapse, while a less massive star ($\sim 35 M_\odot$) progresses from
an O main sequence star to a red supergiant to a Wolf-Rayet star of the
WN class to a supernova collapse.  Significant evidence and theoretical
models support the hypothesis that long-duration \grbs\ are caused by the
collapse of massive stars \citep{wheel00,woo01}, consistent with the
notion that the immediate progenitors of bursts are Wolf-Rayet stars.

Each stage in the evolution of massive stars has a distinct wind
\citep{gml96,glm96}.  The O main sequence star wind is characterized by
a velocity of $\sim 3000$ \kms\ and a mass loss rate of order $10^{-6}$
to $10^{-5} M_\odot {\rm yr}^{-1}$ which lasts for a few million years.  
The Luminous Blue Variable phase has a massive ejection (at a rate of up
to $10^{-3} M_\odot {\rm yr}^{-1}$) over a short time (roughly 10,000
years) with moderate velocities (around 300 \kms).  The red supergiant
wind is characterized by a velocity of 10-100 \kms\ and a mass loss rate
of up to $10^{-4} M_\odot {\rm yr}^{-1}$ which lasts for around 200,000
years.  The Wolf-Rayet wind is characterized by velocities of 1000-3000
\kms\ and a mass loss from $10^{-5}$ to $10^{-4} ~ M_\odot {\rm
yr}^{-1}$ over a lifetime of fractions of a million years.

The interactions between the winds in each successive stage of evolution
will create shells with various properties \citep{gml96,glm96}.  The O
main sequence star will blow a bubble into the surrounding ISM that has a
thin shell and a slow expansion velocity at a distance of order 30 pc.  
The massive ejection of a Luminous Blue Variable will produce a thick
shell that expands with a velocity of around 300 \kms.  The red
supergiant wind will bunch up at its outer edge and expand with a
velocity of 10-100 \kms.  The Wolf-Rayet wind blows a bubble in the
slower and denser wind around it, creating a thin shell that expands with
a velocity of order 500 \kms.  This bubble will overtake the slower shell
in $\sim 10,000$ years and coalesce to form a combined turbulent shell.  
At the time of the core collapse of the star, the inner several parsecs
will consist of the Wolf-Rayet wind expanding at 1000-3000 \kms,
surrounded by a dense and turbulent shell expanding at perhaps 500 \kms\
at a distance of around 10 pc, surrounded by a relatively low density
region evacuated by the O main sequence star wind, all inside a
geometrically thin bubble of material piled up from the ISM.  The winds
of Wolf-Rayet stars are likely to be anisotropic, and the absorption
lines in GRB~021004 were created by only a small geometrical cross
section, presumably along the rotation axis of the progenitor star.

Can this wind structure around a very massive star account for the
observed velocity components in the absorption lines of GRB~021004?  The
blue shifted 3000 \kms\ component would arise from the dense inner wind
driven during the Wolf-Rayet phase.  The O main sequence star wind also
has a velocity of 3000 \kms, but its density has been greatly reduced by
its expansion into space so that it should produce only negligible
absorption. The requisite density in this radius range could also be
provided in principle by the slower wind of a lower mass star with lower
mass loss rate; this case would subsequently require more radiative
acceleration.  If the density in the wind falls off as the inverse-square
of the radial distance from the star, the absorption from the wind will
be dominated by its inner region, say 0.2--0.4 pc.  This means that the
absorption lines would primarily sample a Wolf-Rayet wind emitted over a
brief time interval somewhat before the core collapse.  Over its
lifetime, a Wolf-Rayet wind is roughly constant in velocity, while the
wind velocity across the 0.2--0.4 pc region should be even more precisely
constant.  This could produce a narrow absorption line just as is
observed.  The blue shifted 560 \kms\ absorption component in GRB~021004
could arise from the dense shell created at the outer edge of a
Wolf-Rayet wind.  This shell is expected to be composed of geometrically
thin structures that form a turbulent front width of which is roughly a
tenth of its radius which expands at roughly 500 \kms\
\citep{gml96,glm96}.  The velocity structure within the shell is chaotic
and will lead to some broadening of the resultant absorption lines.  
Thus, expected Wolf-Rayet wind structures can produce the observed 3000
\kms\ and 560 \kms\ components.

What are the expected column densities in the Wolf-Rayet winds?  For a
mass loss rate of $3 \times 10^{-5} M_\odot {\rm yr}^{-1}$, a wind
velocity of 3000 \kms, and an average atomic weight equal to that of
helium, a Wolf-Rayet wind will produce a column density of roughly $3
\times 10^{17} {\rm atoms ~ cm}^{-2}$.  The total column density of the
shell at the outer edge of a Wolf-Rayet wind is uncertain, but it is
likely to be roughly the mass ejected \citep{gml96,glm96} in the red
supergiant phase ($\sim$ 20 $M_\odot$) or the Luminous Blue Variable
phase ($\sim$ 10 $M_{\odot}$) spread over a shell of radius 10 pc.  For
an average atomic mass like that of helium, the total column density will
be of order $2 \times 10^{17}$ atoms cm$^{-2}$ or greater.  The two shell
components thus have comparable column densities, just as the two
absorption lines have comparable equivalent widths.

Are these column densities enough to produce significant absorption
lines?  For the oscillator strengths of \ion{C}{4} and an assumed
velocity dispersion of order 100 \kms, a column density of $\sim 10^{14}
~ {\rm cm}^{-2}$ for \ion{C}{4} atoms in their ground state will produce
an absorption line with an optical depth of unity.  With the total column
density for a component of a Wolf-Rayet wind at around $3 \times 10^{17}$
atoms cm$^{-2}$, the carbon line would produce a significant feature
provided that \ion{C}{4} in the ground state is more than $\sim 0.0003$
of the nucleons in the wind.  A large fraction of the ejecta from
Wolf-Rayet stars will be carbon.  The various ionization states will all
spend almost all the time at their ground level because the decay times
are of order $10^{-8}$ seconds.  The question of calculating the fraction
of carbon that is ionized to any particular degree (by the radiation from
the burst and from the progenitor) is a difficult problem both for the
physics and for the lack of knowing the conditions of the gas (\S 3.2).  
Nevertheless, it is plausible that some significant fraction of the
carbon will be three-times ionized.  For one-tenth of the carbon being
three-times ionized, the total column of atoms capable of producing the
\ion{C}{4} line in the Wolf-Rayet wind will be $\sim 10^{16}$ atoms
cm$^{-2}$.  For such a case, the line will be highly saturated.  This
estimate has many uncertainties, but it is clear that Wolf-Rayet winds
can have sufficient carbon to create the observed absorption lines.

Wolf-Rayet stars often show P-Cygni line profiles, with the emission
component arising from portions of the shell off the line of sight.  If
the wind structure described above exists around GRB~021004, we do not
expect to see P-Cygni profiles.  There are two reasons for this.  The
first is that the \grb\ radiation is strongly beamed by the jet, so most
of the shell off the line of sight is not illuminated.  The second reason
is that a flash of illumination on the shell will produce emission lines
that are delayed by an amount that varies with position in the shell.  
For typical shell sizes, the emission component would be spread out over
many years and will become too faint to be observable.  Thus the lack of
P-Cygni profiles in the high-excitation lines of GRB~021004 is not an
argument against their origin in a Wolf-Rayet wind.

What is the lower limit on the mass of the original star such that it can
still produce the required dense wind?  O-type stars with masses of $\sim
30 M_{\odot}$ can produce dense winds with surrounding shell structures
\citep{kp00}, and hence might account for the observed lines in
GRB~021004.  These stars can have wind velocities up to 3000 \kms\ but
with mass loss rates that might be smaller than allowed by our analysis
of the spectral energy distribution (see \S 4).  The evolution of such
stars would be from an O main sequence star to a red supergiant to a blue
supergiant with a core collapse at some point.  While such a progenitor
cannot be called a Wolf-Rayet star, it is still a massive star surrounded
by shells and hence could be the progenitor of GRB~021004.  Since the
clumpy density profile is so uncertain, the only rigorous lower limit we
can place on this process is that for core collapse itself, that is $\sim
8 M_{\odot}$. For instance, a star that blew up as a red giant might have
a carbon-rich wind and hence have the necessary density to give the
column depth of \ion{C}{4}.  Such a structure would have to rely on
radiatively accelerated clumps of matter at $\sim$ 0.5 pc to produce the
lower velocity \ion{C}{4} features in the absence of a shell generated by
the blue/red wind boundary.

\subsection{The Progenitor is a Massive Star}

We now have two scenarios that can explain the velocity and ionization
structure of the high-excitation absorption lines seen in GRB~021004.  
The first scenario postulates two gas shells or clumps initially nearly
at rest at distances of $\sim 0.2$ pc and $\sim 0.5$ pc that are
radiatively accelerated to velocities of 3000 \kms and 560 \kms,
respectively, as well as ionized by the flash from the burst itself.  
The second scenario proposes that the progenitor is a Wolf-Rayet-like
star with a wind velocity of 3000 \kms\ that is surrounded by a shell
with velocity 560 \kms, the gas of which is ionized by the flux from the
progenitor.  These two scenarios are actually part of a continuum of
similar scenarios where the ionizations and velocities vary from being
caused entirely by the progenitor to being caused entirely by the burst.  
For example, in the second scenario, the original wind might have started
at a lower ionization state at a velocity of 1500 \kms\ and been
accelerated and ionized to the final values by the burst.  In all these
related scenarios, the absorption lines with near zero velocity are
caused by chance clouds along the line of sight as far as 2500 pc from
the burster that have been ionized in part by the burst radiation.

In the first scenario, we have to ask why a \grb\ progenitor would have
two geometrically-thin gas clouds at distances of $\sim 0.2$ pc and $\sim
0.5$ pc?  This is improbable unless there is some causal connection
between the clouds and the burster.  The presence of two
geometrically-thin clouds along the line of sight could be caused by two
shells around the progenitor or by some number of clumps of gas around
the progenitor.  In the second scenario, we also have shells of gas
centered on the progenitor.

We conclude that the progenitor had shells or clumps of gas centered on
it and associated with it.  Models where the \grb\ arises from the
collision of two compact objects in close orbit are substantially less
likely because these have no ready supply of gas close to the system.  
The presence of shells is an unusual situation for stars, as few have
multiple shells, although clumps are more likely.  The only stellar
objects with multiple shells that have any realistic hope of being
associated with \grbs\ are supernovae and massive stars.  Any shell from
a supernova would have to be pre-existing (as in the Supranova model of
Vietri \& Stella, 1998)  since the afterglow jet would be far outside any
normal supernova ejecta if the supernova and burst exploded
simultaneously.  \citet{salamanca02} point out there are difficulties
with producing narrow lines with the observed equivalent width from an
old supernova remnant.  The narrowness of the absorption lines would
require that the ejecta has already cooled and condensed into filaments.  
The question is then the likelihood of getting multiple filaments of
greatly different velocity on our line of sight and the column density of
any such filaments.

Ideally, the absorption lines from the $z \cong 2.293$ system can tell us
about the abundances of carbon, silicon, hydrogen, and nitrogen in the
innermost shell ejected by the progenitor; and this can then tell us
about the progenitor itself.  For example, if the composition is that of
a Wolf-Rayet shell of a particular type (perhaps with some swept up ISM
material), then we would even get a good idea of the mass of the
progenitor.  Unfortunately for this program, we have few lines from the
$z \cong 2.293$ system (see the many lines for GRB~010222,
\citet{mirabal02a}). Also, the likelihood that our unresolved lines are
saturated implies that the measured equivalent widths will be difficult
to convert to abundances. Nevertheless, a full analysis based on our data
or the data of, for example, \citet{savaglio02}, might place useful
limits on the abundances of the shell and its possible origin.

An example of such an analysis can be made with respect to the
carbon-to-nitrogen abundance.  Both atoms have similar ionization
potentials and should have somewhat similar ionization fractions into
either \ion{C}{4} or \ion{N}{5}, the oscillator strengths of the two
resonance lines are similar, and the cosmic abundance of nitrogen is down
from that of carbon by a factor of three.  So simplistically, we would
expect the \ion{C}{4} line to have an equivalent width that is roughly
three times deeper than that of \ion{N}{5}, whereas (see Table 1) we do
not see any nitrogen lines and the limit on the equivalent widths is a
factor 4.6.  With the likelihood that the \ion{C}{4} line has a saturated
core, the implication is that nitrogen is significantly under-abundant
compared to carbon.  This would appear to rule out WN type Wolf-Rayet
stars.  The lower mass Wolf-Rayet stars undergo core collapse while the
wind is still nitrogen rich \citep{glm96}, whereas the higher mass
Wolf-Rayet stars have core collapse after the Wolf-Rayet has turned from
a WN to a WC type \citep{gml96}.  Thus, the lack of a \ion{N}{5} line in
the spectrum of GRB~021004 appears to imply that the burst progenitor
must be amongst the more massive of Wolf-Rayet stars.  Alternatively, the
progenitor might be a star for which the original mass was too low ($<$
$30 ~ M_{\odot}$) to allow it to become a WN-type star, yet it would
still have fast and dense winds that are not nitrogen-rich.

The lack of molecular hydrogen absorption can be used to place further
limits on the density of the gas near the burster.  In the model of
\citet{draine02}, the column density of excited $H_2$, $N(H_2^*)$, arises
from an ionization/dissociation front the dynamics of which imply that
there are natural values of the column depth.  For example, an increase
in the burst fluence would produce a similar front (only deeper into the
molecular cloud) with a similar $N(H_2^*)$, while an increase in the
molecular density would produce a front with a narrower width yet with a
similar $N(H_2^*)$. The only way to produce a smaller column density is
to have UV fluxes that are not sufficient to establish a proper
ionization/dissociation front in the molecular hydrogen.  This could
arise if the molecular cloud were at a great distance (say, 1 kpc) from
the burster or if the burster were near the edge of the molecular cloud.  
Draine calculates that a typical burst will destroy molecular hydrogen in
gas with a hydrogen column density of $5 \times 10^{22} {\rm cm}^{-2}$
(corresponding to $A_V = 20$ or so).  For densities from 100--10000 ${\rm
cm}^{-3}$, the burst will destroy the molecular hydrogen that could serve
as an absorber if it is within roughly 2--200 pc of the edge of the
cloud.  For the general case of the burst erupting well inside a
molecular cloud, the $N(H_2^*)$ will be the value used to calculate the
template (see Section 2), and this is rejected by our spectrum.  We
conclude that the lack of molecular hydrogen absorption lines in our
spectra indicate that the \grb\ is not inside a molecular hydrogen cloud
unless it is within $\sim$2-200 pc from its edge.

\section{The Spectral Energy Distribution}

We have constructed broad-band fluxes from the radio to the x-ray for
several epochs (see Figure 3) with the data collected from the GCN
notices. The major features of the radio emission are:  $i)$ at 0.5--3.5
days the 15 GHz light-curve rises as $t^{0.4\pm0.3}$, $ii)$ at 1.5 days
the 15--86 GHz spectrum is $F_\nu \propto \nu^{0.6\pm0.2}$, appearing
flat from 86 to 230 GHz, and $iii)$ at 5.7 days the 1.4--8.5 GHz spectrum
is $F_\nu \propto \nu^{1.0 \pm 1.6}$. The last two features indicate that
the synchrotron self-absorption frequency $\nu_a$ is not significantly
above 15 GHz at 1.5 days and not much below 8 GHz at 5.7 days, which
would suggest a wind-like medium (for which $\nu_a \propto t^{-3/5}$)
instead of a homogeneous environment (for which $\nu_a$ is constant).
However, with only a few measurements and the possibility of significant
interstellar scintillation affecting the lower frequency (under 15 GHz)
observations, it is premature to favor a wind-like medium based on the
apparent decrease of $\nu_a$ over a factor of 4 in time. The rise of the
15 GHz emission is consistent with both the expectations from a
homogeneous medium (where $F_\nu \propto t^{1/2}$ irrespective of the
location of $\nu_a$) and a behavior intermediate to the asymptotic
$F_{\nu \ll \nu_a} \propto t$ and $F_{\nu \gg \nu_a} = const$ expected
for a wind-like medium.

Within the framework of relativistic fireballs, consistency between the
decay rate of the optical emission ($F_\nu \propto t^{-\alpha_O}$ with
$\alpha_O \simeq 1.0$) and spectral slope ($F_\nu \propto \nu^{-\beta_O}$
with $\beta_O \simeq 1.0$) requires that the cooling frequency $\nu_c$ is
below or within the optical range, as in this case the expected
relationship ($\beta_O=(2\alpha_O+1)/3$, irrespective of the type of
external medium)  is satisfied. Furthermore, that $\nu_c$ is below
optical is also consistent with the equality of the optical and X-ray
decay indices ($\alpha_X = 1.0 \pm 0.2$; Sako \& Harrison 2002) and
spectral slopes ($\beta_X = 1.1 \pm 0.1$). We note, however that the
difference between the observed optical and X-ray decay indices,
$\alpha_X - \alpha_O = 0.2$, is also consistent (even if only marginally)
with the theoretically expected value $\alpha_X - \alpha_O = 0.25$ in the
case where $\nu_c$ is above the optical domain but below X-rays, and that
the intrinsic optical spectral slope $\beta_o$ can be smaller than
observed if there is a significant dust reddening within the host galaxy.
Therefore, the cooling frequency could be above the optical domain, in
which case one expects that $\beta_o = 2\alpha_O/3 = 2/3$ for a
homogeneous medium, and $\beta_o = (2\alpha_O - 1)/3 = 1/3$ for a
wind-like medium.  Furthermore it is also expected that $\beta_o =
\beta_X - 0.5 = 0.6 \pm 0.1$, which favors a homogeneous medium when
compared with the above $\beta_o$ inferred from the optical decay.

Concluding, the currently available radio and optical data for GRB~021004
do not constrain sufficiently the type of external medium, homogeneous or
wind-like. However, the average density of the medium or the mass-loss
rate--to--wind speed ratio can be determined if the cooling frequency
lies within the optical range at about 2 days, as suggested by Matheson
et al 2002.  Together with the optical measurements, the 15, 86 and 230
GHz fluxes reported by Pooley (2002) and Bremer \& Castro-Tirado (2002)
indicate that, at 1.5 days, the peak of the spectrum of GRB~021004 is
$\sim 3$ mJy, at a frequency between $10^{11}$ and $10^{12}$ Hz. Adding
that $\nu_a \simeq 10$ GHz at a few days, these four constraints on the
synchrotron spectrum are sufficient to determine the external medium
density, fireball isotropic-equivalent energy, and fractional energies in
electrons and magnetic fields. We obtain an external density in the range
$10^{0.5}- 10^{2.5}\; {\rm cm^{-3}}$ for a homogeneous medium, consistent
with the values found by Panaitescu \& Kumar (2002) for other GRB
afterglows. For a wind-like medium, we find that density of the wind
corresponds to that resulting from a mass-loss rate--to--wind speed ratio
in the range 1 to several times $10^{-5} (M_\odot {\rm yr}^{-1})/(1000$
\kms), within the range expected for WR stars.

\section{The Light Curve}

We have constructed light curves and color curves for GRB~021004 from data
published in the GCN notices (see Figure 4).  Overall, the light curve
behaves approximately as a $t^{-1}$ power law which is normal for many
afterglows.  However, there are many significant bumps and wiggles about
any simple single power law.  To help visualize these bumps, the middle
panel of Figure 4 shows the R-band light curve after a $t^{-1}$ power law
has been subtracted, and we see three bumps with peaks at 0.14, 1.1, and
4.0 days after the burst.  The V-R color is plotted in the lower panel of
Figure 4.  The color appears to remain essentially constant from 0.2 to 10
days.

Afterglow models suggest that light curves should consist of broken power
laws, so there is an urge to try to interpret Figure 4 in terms of broken
power laws.  Perhaps the basic behavior is $t^{-1}$ with bumps and some
separate mechanism for the dimness at the earliest times.  
Alternatively, perhaps the basic behavior is a $t^{-0.7}$ power law up
until a break at 3 days to a $t^{-1.5}$ power law with a particularly
strong bump from 0.1--0.7 days.  For a third alternative, perhaps the
basic behavior is a $t^{-1.5}$ decline over the entire time range except
that some mechanism inserts more energy three times during the decline
(at 0.14, 1.1, and 4.0 days).

In general, \grb\ afterglows have simple power law declines in
brightness, often with a break ascribed to collimation of the outflow.  
Nevertheless, many \grbs\ have bumps in their afterglows.  For example,
the optical brightness of the afterglow of GRB~970508 increased by 1.5
magnitudes 1--2 days after the \grb\ \citep{garcia98}.  GRB~970508 was
seen by the BeppoSAX satellite to have a significant brightening in its
x-ray light curve starting at 16 hours after the burst \citep{piro99}.  
GRB~000301c displayed an achromatic bump that peaked around 4 days after
the burst that has been ascribed to gravitational microlensing by
\citet{gls00}.  In addition, three \grbs\ with bumps around 30 days after
the burst have been ascribed to an underlying supernova (Reichart 1999,
Bloom et al. 1999; 2002; but see Dermer 2002)  or dust echos (Reichart
2001).

The presence of bumps in otherwise-smooth afterglow light curves has
previously been attributed to microlensing, dust echos and supernovae.  
Now we have a burst with three bumps, and they cannot all be due to
microlensing, dust echos, or supernovae, so we have a proof that
afterglows can makes bumps on various time scales for other reasons.  
Thus, the mere presence of a single bump can only provide poor support
for any particular proposed cause unless other possibilities can be
excluded by some means.  For example, the red color of a bump (compared
to the underlying afterglow) cannot be taken as substantial evidence of a
supernova origin until the cause and color-distribution of other classes
of bumps are known.

There are two generic scenarios (in addition to microlensing, dust echos,
and supernovae) for producing variability in the afterglow emission: an
inhomogeneous circumburst medium and non-uniform properties of the \grb\
ejecta (see also Nakar, Piran \& Granot 2003). Inhomogeneities in the
external gas could be either in the form of clumps of denser material
\citep{wangloeb00}, which enhance the dissipation of the kinetic energy
of the \grb\ remnant and, implicitly, the afterglow brightness, or in the
form of a stratified medium, consisting of shells of various densities
\citep{lazzati02}, perhaps produced by fluctuations in the wind blown by
the \grb\ progenitor not long before the release of the \grb\ ejecta.  
Non-uniformities in the \grb\ ejecta could be manifest either as a range
of initial Lorentz factors within an impulsive ejection or as a
non-isotropic distribution of the ejecta energy per solid angle. In the
former case, the slower part of the outflow catches up with the faster
part as the latter is decelerated, enhancing the afterglow kinetic energy
and, hence, its brightness. In the latter case, the regions of higher
angular energy density are bright spots on the ejecta surface, moving
slightly off the direction toward the observer, so that fluctuations in
the afterglow emission are produced as the spots decelerate and the
widening cone of their relativistically beamed emission starts to include
the direction toward the observer.  The variable-Lorentz-factor and
anisotropic outflow scenarios have been used to explain the brightening
of the afterglow of GRB~970508 by \citet{pmr98} and \citet{pk02},
repectively.

An energy input that is extended in time prior to the brightening at 0.14
days could explain the shallower decline prior to 0.14 days, while an
inhomogeneous shell with a density decreasing with radius
\citep{lazzati02} could account for the steeper light-curve decay after
0.1 days in the stratified medium scenario. The faster decline after a
brightening is consistent with the expectations from scenarios involving
dense clumps or bright spots, where the contribution of the clump or spot
to the total afterglow emission decreases in time as the fireball
decelerates, and more of the uniform ejecta surface becomes visible for
the observer.

In the latter two scenarios where the afterglow emission from clumps or
spots declines with time, the afterglow light curve should eventually
recover its initial decay rate. In this context, the observed light curve
is consistent with the proposal that the $t^{-0.7}$ fall-off seen in the
first 15 minutes reappeared at 0.7 days, just about the time the spectra
reported here were obtained, and lasted until 1--3 days after the burst,
when the light curve steepened to $t^{-1.0}$.  If this interpretation is
correct, GRB~021004 would display the shallowest light-curve break ever
observed.  The change in the break index would be $\simeq 0.3$,
suggesting that the break is caused by the passage of the cooling break
through the optical domain \citep{pk01}.  The existence of a break in the
optical spectrum at 1--3 days is supported by the spectroscopic
observations by \citet{matheson02} and this paper (see \S 2), that the
blue part of the optical spectrum reddens between $< 1$ and $> 2.8$ days.
This interpretation is consistent with a break frequency that decreases
in time, implying that the external medium is homogeneous rather than
wind-like.  In the case of a wind, the cooling frequency should increase
at least as fast as $t^{1/2}$ \citep{pk01}.  We note, however, that the
break in the optical spectrum seems too sharp in wavelength to be due to
the cooling break (across which the slope of the power-law spectrum
changes by 0.5), and that the spectral changes may not sufficiently
affect the R-band light-curve at a few days to explain the R-band light
curve behavior.

The radial distance of the jet from the center of the progenitor as a
function of the observer's time should be correlated with the location of
clumps and the epoch of the light curve bumps.  This can be modelled
given both the broad-band spectrum and the light curve by the methods
presented in \citet{pk01,pk02}. Unfortunately, only preliminary input
numbers are available (for example, no jet break has yet been definitely
timed) and details of the external medium are not known (for example,
whether or not the average density falls off like a wind).  
Nevertheless, it is possible to give estimates that are useful for
producing a reasonable picture.  For this, we adopt the isotropic
equivalent energy of the fireball equal to $5.6 \times 10^{52}$ erg
\citep{malesani02}.  If the jet is expanding into an external medium with
a wind-like structure, then we found in the previous section that the
mass-loss-rate to wind-speed ratio is roughly $2 \times 10^{-5} M_\odot$
per 1000 \kms.  In this case, the radius of the jet will be $t^{0.5}
\times 0.06$ pc, with $t$ measured in days.  If the jet is expanding into
a homogenous medium of constant density, then we found in the previous
section that the density is roughly 30 cm$^{-3}$.  In this case, the
radius of the jet will be $t^{0.25} \times 0.07$ pc.

If the light curve bumps are due to density enhancements in the
surrounding medium, then we can estimate the approximate radial distances
from the burst and over-densities for each clump.  For peaks in the light
curve at 0.14, 1.1, and 4.0 days, we find approximate radii for the
density clumps of 0.022, 0.063, and 0.12 pc, respectively, for a
wind-like medium.  We get similar results for expansion into a homogenous
medium with parameters as derived above.  The excess light in the light
curve bumps will scale roughly as the square root of the over-density
compared to the ``background" density, for the case where the clump is
relatively large compared to the visible region (as for a shell).  From
Figure 4, the extra light in each of the bumps is roughly 0.6, 0.3, and
perhaps 0.3 mag; this corresponds to over-densities of roughly 50\%,
10\%, and 10\%.  If the bumps are due to clumps, the broad-band spectral
slopes should remain the same and thus the colors would not change across
the bump, as observed.

At some time the afterglow jet will collide with the gas that is causing
the $z \cong 2.293$ absorption.  When this happens, we would expect that
the afterglow will brighten significantly if the gas is in a relatively
narrow shell.  We also expect that the associated absorption lines will
disappear.  We note that our late HET spectrum still showed the $z \cong
2.293$ \ion{C}{4} line with moderately low significance at a time of 4.84
days after the burst when the jet would have reached $\sim 0.13$ pc.  In
principle, we can predict the date at which this bump in the afterglow
decline should occur.  In practice, such a prediction will have
substantial uncertainties, primarily in the exact radius of the shell and
the density structure of the medium inside the shell.  In the case where
a shell of material has been radiatively accelerated to 3000 \kms, the
shell will have a radius not much larger than 0.1 pc and it should be
over-run sometime between roughly 2--40 days after the burst. In the case
where the absorption is due to a Wolf-Rayet wind that extends out to
several parsecs, the 3000 \kms\ component will be completely eliminated
only after many years.  The absorption will slowly diminish during the
first few months after the burst as the jet catches up with the densest
inner part of the wind.  In summary, the absorbing gas might be
eliminated during the early days after the burst (an event that could be
directly observable) or it might take many years to be eliminated (and be
totally unobservable).

\section{Conclusions}

The GRB~021004 spectrum displays high-ionization and high-velocity
absorption lines.  We argue that these lines are likely to have arisen
from shells of gas around the progenitor star, and further, that these
shells plausibly come from a progenitor that was descended from a massive
star.  The gas in these shells was partially ionized and accelerated by
the photons from the burst.

We have used the optical spectrum, the broad-band spectrum, and the light
curve to identify six components in the medium external to the \grb.  
The $z \cong 2.293$ absorption lines arise from a shell at a radial
distance of less than a parsec that is moving at a velocity of 3000 \kms
after being ionized and accelerated by the burst radiation.  We identify
this component as being the wind from the progenitor star, perhaps a
Wolf-Rayet star with a mass loss rate of $\sim 6 \times 10^{-5} M_\odot
{\rm yr}^{-1}$ or a lower mass star wind accelerated by the bust.  This
same component provides the external medium with which the jet is
colliding to produce the afterglow light.  This inner component is
probably significantly clumped in order to prevent complete ionization
and to facilitate recombination, thus allowing a significant column depth
of \ion{C}{4}.  The second component is associated with the absorption
lines moving 560 \kms\ with respect to the host galaxy velocity.  We
identify this component as arising from a shell of material plowed up by
the Wolf-Rayet-like wind at a radius of order 10 parsecs.  
Alternatively, this feature could be a shell or clump at $\sim$ 0.5 pc
accelerated by the burst. The third component gives rise to the
low-velocity high-excitation absorption lines which must come from a
distance between tens and hundreds of parsecs from the burster.  This
component might be from the shell created by the O main sequence star
wind or it might be from chance ISM clouds along the line of sight,
perhaps part of the same star forming region that gave birth to the \grb\
progenitor.  The fourth, fifth, and sixth components are three density
enhancements that occur in the Wolf-Rayet-like wind at radial distances
of 0.022, 0.063, and 0.12 pc so as to produce the three bumps in the
light curve. These clumps could be related to the clumping we deduce for
component 1, that is, all could be associated with knots in the
progenitor wind.

The results summarized in the previous paragraph offer a plausible
description of the density structure near this burst. What we see is
a variety of density enhancements along the line of sight, with the
evidence coming from bumps in the light curve decline and absorption
lines in the spectrum.  This picture is apparently not universal for
\grbs, since only two out of nine bursts with well-observed
afterglow spectra show both high-excitation and high-velocity lines.  
The other bursts must have some mechanism for suppressing similar
absorption lines.

A full analysis is needed for the $z \sim 2.3$ lines of GRB~021004, with
the possibility of determining the abundance of the progenitor's shell
and hence to obtain information about the progenitor itself.  As a start
to this program, we have shown that the nitrogen in the shell is
under-abundant with respect to the carbon, and this demonstrates that the
progenitor was not a WN-type Wolf-Rayet star and further that the
progenitor was either one of the more massive Wolf-Rayet stars or a
massive star that is below that required to make a WN-type star.  A
further program for our community is to obtain high dispersion
spectroscopy so as to be able to measure fine velocity structures and
faint lines in the afterglow spectra of many bursts.

The most important result from our data and analysis is that a plausible
way to create the high-excitation and high-velocity absorption lines is
to have shells of gas surrounding the progenitor, and that a realistic
way to achieve this is for the progenitor to be a star descended from a
massive O-type main sequence star.  This strongly suggested connection to
a massive star is much more direct than the moderate statistical
association of \grbs\ with parts of galaxies undergoing star formation
and is much more convincing than attributing poorly observed bumps in
afterglow light curves to some associated supernova event.  As such, we
believe that the absorption lines in the GRB~021004 spectra provide 
the
first strong and direct argument for associating a normal \grb\ with a
massive progenitor star surrounded by shells.

We thank Bruce Draine, Bev Wills, Greg Shields, and D. J. Hillier for
their discussion on a wide range of topics.  This work was supported in
part by NASA Grant NAG59302 and NSF Grant AST-0098644.  The Marcario Low
Resolution Spectrograph is a joint project of the Hobby - Eberly Telescope
partnership and the Instituto de Astronomia de la Universidad Nacional
Autonoma de Mexico. The Hobby - Eberly Telescope is operated by McDonald
Observatory on behalf of The University of Texas at Austin, the
Pennsylvania State University, Stanford University,
Ludwig-Maximilians-Universit\"at M\"unchen, and 
Georg-August-Universit\"at
G\"ottingen.

\clearpage

\begin{deluxetable}{ccccc}
\tabletypesize{\scriptsize}
\tablecaption{Line list for $z \sim 2.3$ component.
\label{tbl1}}
\tablewidth{0pt}
\tablehead{
\colhead{$\lambda_{obs} (\AA)$}   &
\colhead{Line Identification} &
\colhead{$\lambda_{rest} (\AA)$} &
\colhead{$EW_{obs} / (1+z) (\AA)$\tablenotemark{a}}  &
\colhead{$z$\tablenotemark{b}} 
}
\startdata
$4587.2 \pm 3.6$ & Si IV & 1393.8 &  $ 0.60 \pm 0.24 $ & 2.291 
\\
$4628.8 \pm 1.3$ & Si IV\tablenotemark{c} & 1393.8 &  $  2.44 
\pm 0.30 $  & 2.321 \\
$4628.8 \pm 1.3$ & Si IV\tablenotemark{c} & 1402.8 &  $  2.44 
\pm 0.30 $  & 2.299 \\
$4665.9 \pm 1.8$ & Si IV & 1402.8 &  $ 1.17 \pm 0.24 $  &2.326 
\\
$5106.2 \pm 0.9$ & C IV  & 1548.2 \& 1550.8 &  $ 1.39 \pm 0.12 $
 & 2.296 \\
$5149.9 \pm 0.4$ & C IV  & 1548.2 \& 1550.8 &  $ 5.42 \pm 0.18 $
 & 2.325 \\
$5559.9 \pm 3.5$ & Al II & 1670.8 &  $ 0.66 \pm 
0.33\tablenotemark{d} $ & 2.328  \\
$\sim 4095     $ & N V  & 1238.8 \& 1242.8 &  $ \le 0.30 $  &
$ \sim 2.3$  \enddata
    
\tablenotetext{a}{The observed equivalent width measured in \AA ngstroms 
corrected to the rest frame of the burst.}

\tablenotetext{b}{The uncertainty in red shift $z$ is 0.001 -- 0.002.}

\tablenotetext{c}{The 4628.8 \AA\ line is a blend of the two Si IV 
doublet lines from the two red shift components.}

\tablenotetext{d}{This line is close to the bright night sky line at 
5575 \AA, so there might be significant systematic uncertainties in the 
equivalent width due to imperfect night sky subtraction.}

\end{deluxetable}

\clearpage
\begin{figure} 
%\columnwidth=0.75\columnwidth 
%\plotone{f1.eps}
\includegraphics[angle=-90,scale=0.6]{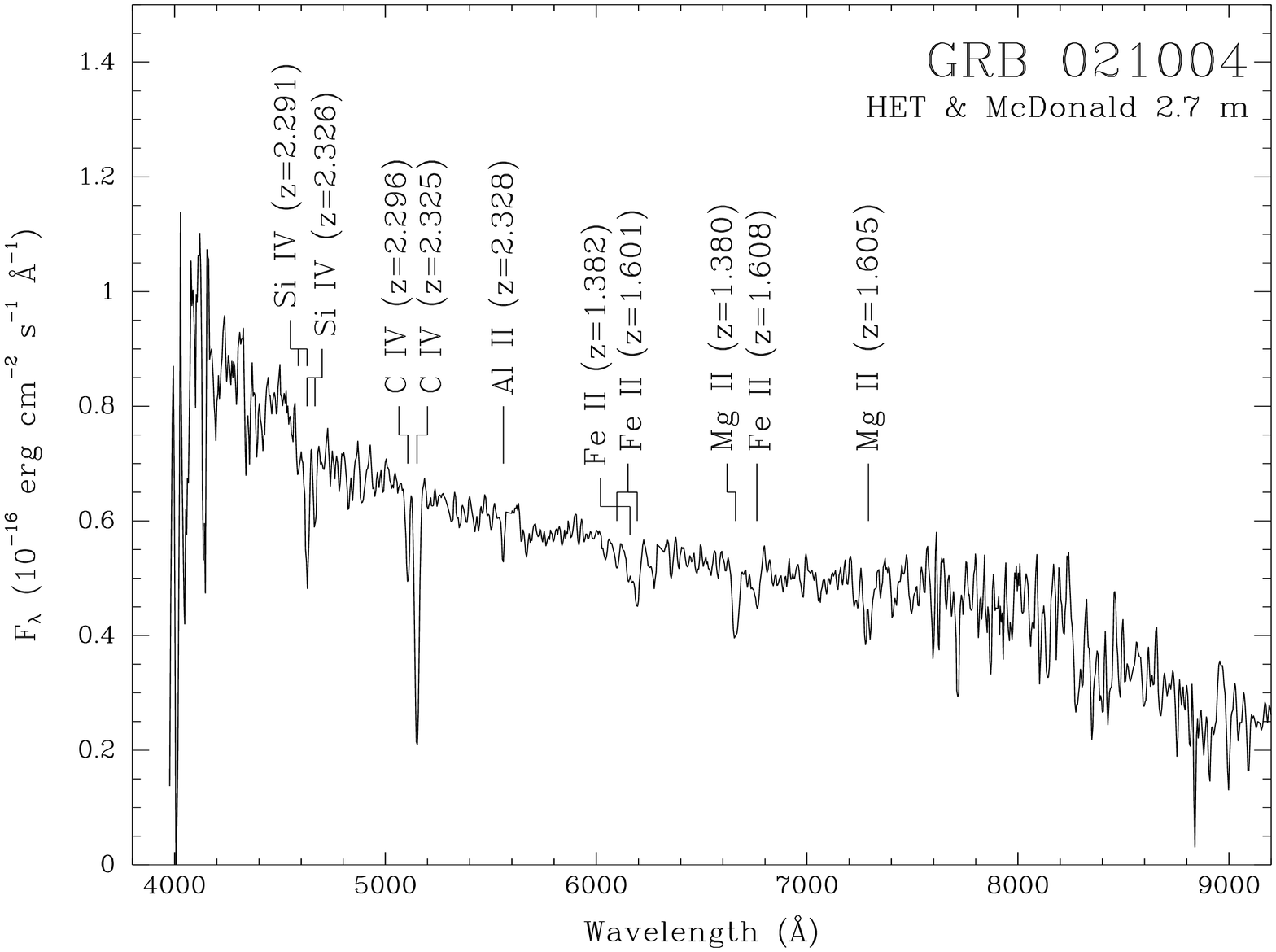}
%\columnwidth=1.33\columnwidth 
\caption{Combined spectrum for the first night.  
The absorption lines longward of 5800 \AA\ are due to intervening
galaxies at red shifts of 1.38 and 1.60.  The primary lines of interest
are the C IV lines from 5106--5150 \AA\ and the Si IV doublet
from 4587--4666 \AA.  Key points in our analysis are that these are highly
ionized lines and that there are multiple velocity components which span a
velocity of 3000 \kms. We identify the lines with red shift of 
around 2.293 as coming
from a shell around the progenitor at a distance of 
less than a parsec.}
\end{figure}

\clearpage 
\begin{figure} 
%\columnwidth=0.5\columnwidth 
%\plotone{f2.eps}
\includegraphics[angle=-90,scale=0.5]{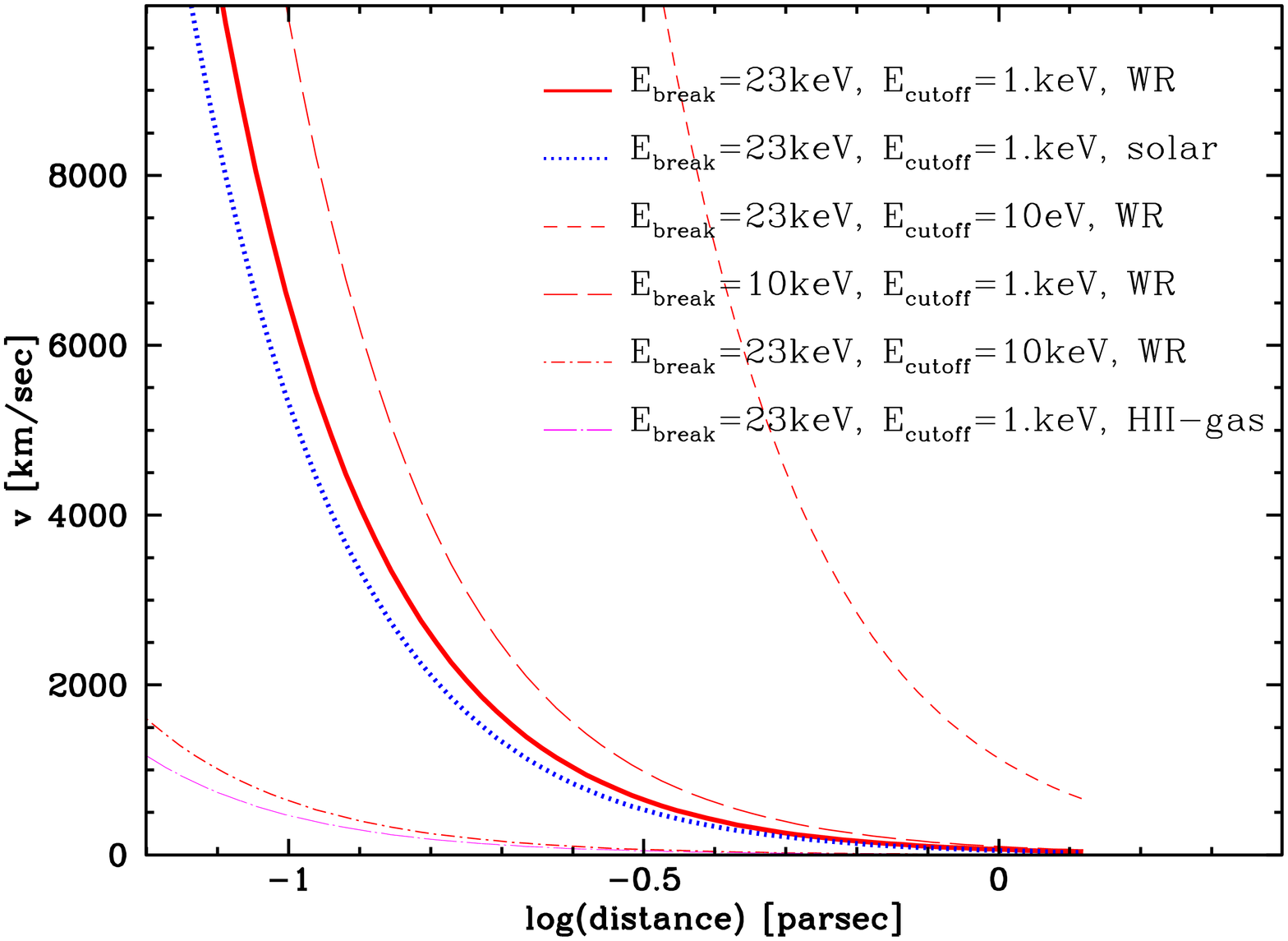}
%\columnwidth=2\columnwidth 
\caption{ 
The terminal velocity induced by GRB~021004 radiation as a function of
distance for three abundances (normal solar abundance, a Wolf-Rayet
abundance with little hydrogen and enriched CNO elements, and a pure
\ion{H}{2} gas) and two values of $E_{break}$ (see text), 10 keV and 23
keV in the frame of the host galaxy (3 keV and 7 keV in the Earth frame).  
The low energy cutoff in the spectrum, $E_{cutoff}$, (0.01, 1.0, and 10
keV in the host galaxy frame) below which we assume there is zero flux
was also varied.  The thick solid line represents the case (WR
composition, the highest break energy, and flux continuing to 1 keV) that
we think is most applicable to GRB~021004.  for which gas at a distance
of around 0.2 pc could be accelerated from rest to 3000 \kms\ by the
burst. The derived radial distances depend little on the gas composition.
The two lowest curves correspond to cases where
the low energy photons do not interact with the gas by bound-free
collisions, demonstrating that most of the acceleration arises from
bound-free interactions with low energy photons.  The case with
$E_{cutoff} = 0.01$ keV is an extreme situation that we do not think is
likely since even a small amount of gas will produce much higher cutoff
energies.
\label{fig2}} 
\end{figure}

\clearpage
\begin{figure}
\columnwidth=0.75\columnwidth
\plotone{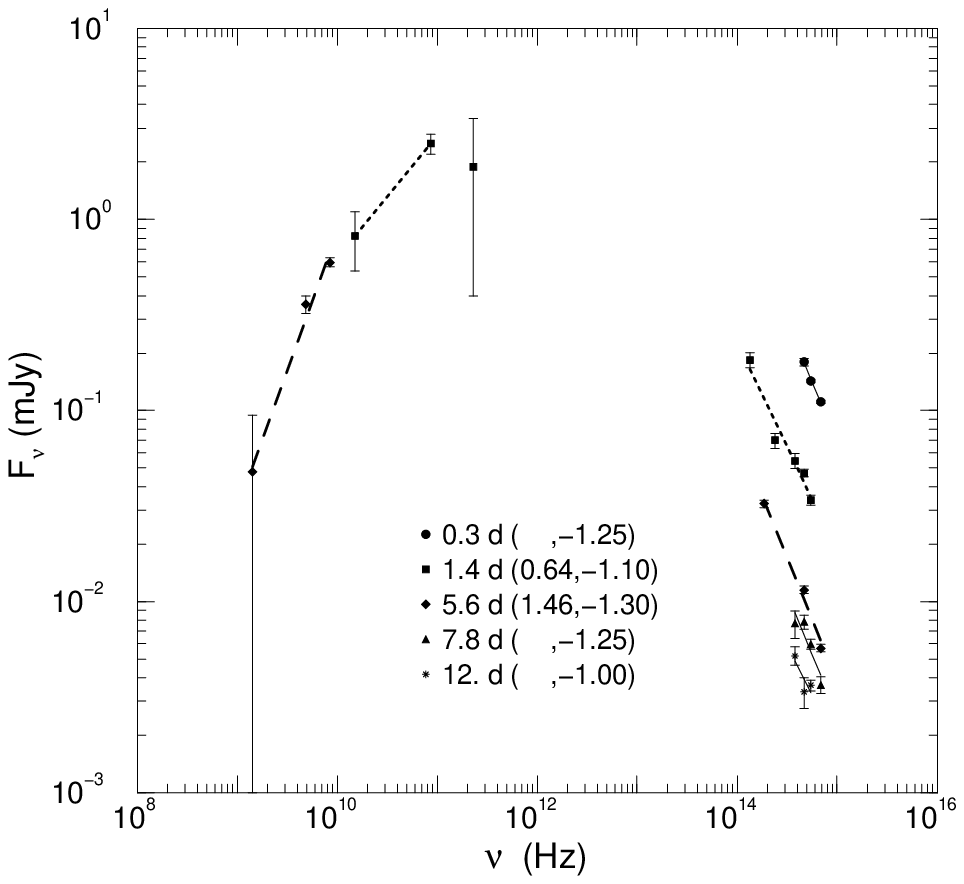}
\columnwidth=1.333\columnwidth
\caption{
Broadband spectrum of GRB~021004 at several epochs.
These composite spectra were constructed from measures given in the GCN 
notices 
\citep{bal02,sahu02,pooley02,bct02,bar02a,berg02,dip02,ste02,mal02b,wil02}.  
The optical light is not corrected for extinction in our Milky 
Way galaxy (which should be moderately low due to the burst's relatively 
high galactic latitude).  In the plot legend, each line indicates the 
type of symbol, the epoch for which the observations were obtained in 
days, while in parentheses the spectral slopes for radio and optical 
are given.  For three epochs, no radio data is available, so the radio 
spectral slope is left blank.
\label{fig3}}
\end{figure}

\clearpage
\begin{figure}
\columnwidth=0.75\columnwidth
\plotone{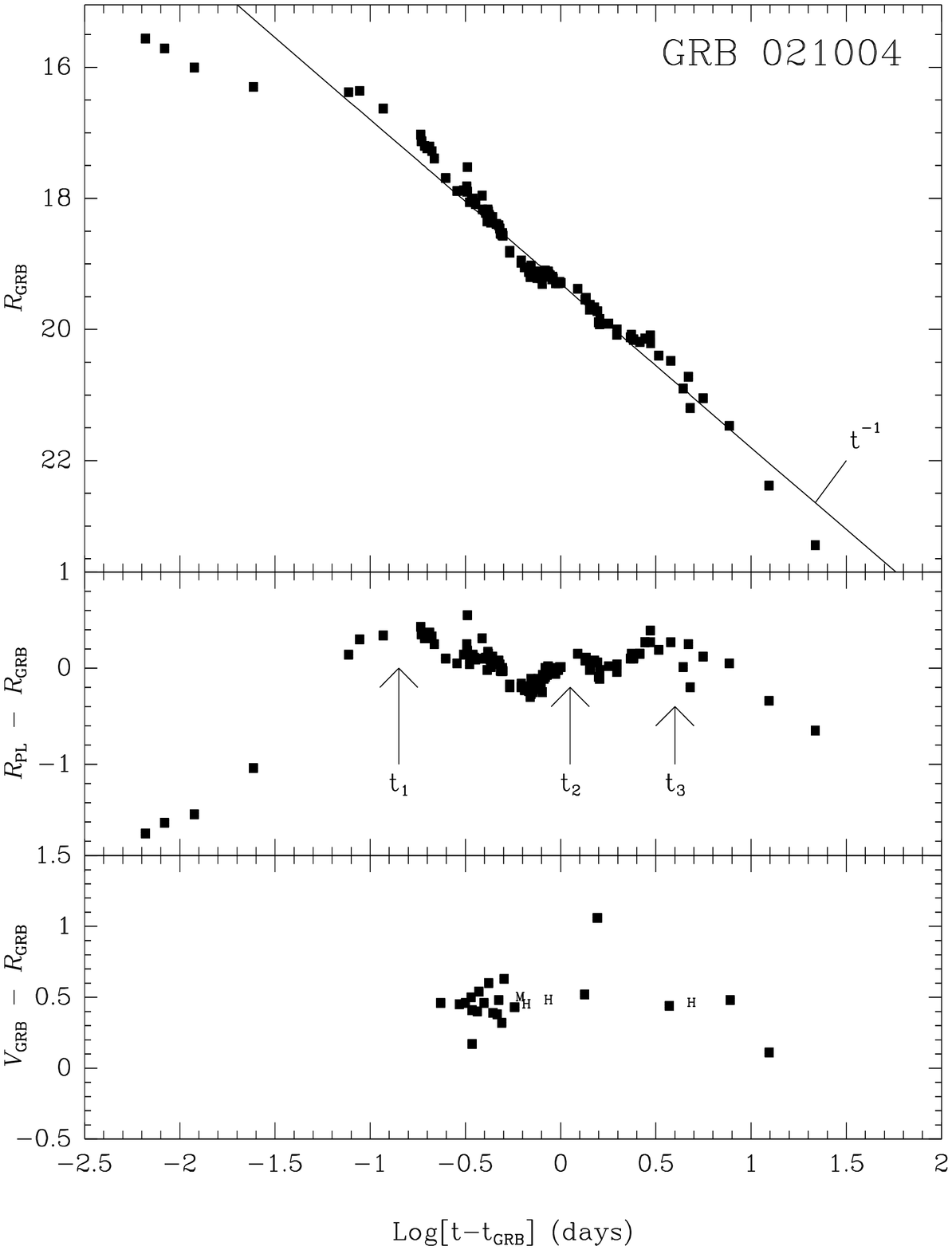}
\columnwidth=1.333\columnwidth
\caption{
Light curve and color curve for GRB~021004.
The top panel displays all R-band magnitudes reported to date in the GCN
notices
\citep{
fox02a,uem02,oks02a,wei02,win02,zha02, hal02b, bal02,coo02, hal02,
bers02,sahu02,oks02b,hal02b,matsumoto02,cov02,hol02c,sta02,mir02c,mas02,
bar02a,malesani02,klo02a,klo02b,dip02,mirabal02b,cov02b,ste02,lin02a,lin02b,
mal02b,wil02,bar02b,gar02}.  The magnitudes have all been reduced to the
system reported by Henden (2002).  For comparison, a $t^{-1}$ power law
is shown.  The middle panel shows the same data with the $t^{-1}$ power
law subtracted.  The important point is that the brightness displays
three peaks or bumps, centered at times of roughly 0.14, 1.1, and 4.0
days after the burst.  The bottom panel shows the V-R color.  Most of the
V magnitudes reported in the GCN are from one single interval (0.3--0.6
days after the burst).  Our V-R colors from synthetic photometry of the
HET and McDonald 2.7m spectra are marked as `H' and `M' respectively.  
We see that the V-R color does not change significantly over the interval
from 0.2--10 days.
\label{fig4}}
\end{figure}


\begin{thebibliography}{}

\bibitem[Anders \& Grevesse(1989)]{ag89}
Anders, E. \& Grevesse, N. 1989, Geochim. Cosmochim. Acta, 53, 197

\bibitem[Balman et al.(2002)]{bal02}
Balman, S. et al. 2002, GCN Circ. 1580

\bibitem[Barsukova et al.(2002a)]{bar02a}
Barsukova, E. A. et al. 2002a, GCN Circ. 1606

\bibitem[Barsukova et al.(2002b)]{bar02b}
Barsukova. et al. 2002b, GCN Circ. 1654

\bibitem[Barth et al.(2002)]{barth02}
Barth, A. J. et al. 2002, astro-ph/0212554

\bibitem[Berger et al.(2002)]{berg02}
Berger, E. et al. 2002, GCN Circ. 1613

\bibitem[Bersier et al.(2002)]{bers02}
Bersier, D., Winn, J., Stanek, K. Z., \& Garnavich, P. 2002, GCN 
Circ. 1586

\bibitem[Bloom et al.(1999)]{bloom99}
Bloom, J. S. et al. 1999, Nature, 401, 453

\bibitem[Bloom et al.(2002)]{bloom02}
Bloom, J. S. et al. 2002, \apj, 572, L45

\bibitem[Bremer \& Castro-Tirado(2002)]{bct02}
Bremer, M. \& Castro-Tirado, A. 2002, GCN 1590

\bibitem[Castander et al.(2002)]{castander02}
Castander, F.~J. et al. 2002, GCN Circ. 1599

\bibitem[Chornock \& Filippenko(2002)]{chorfil02}
Chornock, R. \& Filippenko, A. V. 2002, GCN Circ. 1605

\bibitem[Cool et al.(2002)]{coo02}
Cool, R. J. et al. 2002, GCN Circ. 1584

\bibitem[Covino et al.(2002a)]{cov02}
Covino, S. et al. 2002a, GCN Circ. 1595

\bibitem[Covino et al.(2002b)]{cov02b}
Covino, S. et al. 2002b, GCN Circ. 1622

\bibitem[Dermer(2002)]{dermer02}
Dermer, C. D. 2002, astro-ph/0204037

\bibitem[Di Paola et al.(2002)]{dip02}
Di Paola, A. et al. 2002, GCN Circ. 1616

\bibitem[Djorgovski et al.(2002)]{djorg02}
Djorgovski, S. G. et al. 2002, GCN Circ. 1620

\bibitem[Draine \& Hao(2002)]{draine02}
Draine, B. T. \& Hao, L. 2002, \apj, 569, 780

\bibitem[Eracleous et al.(2002)]{erac02}
Eracleous, M. et al. 2002, GCN Circ. 1579

\bibitem[Fox et al.(2002a)]{fox02a}
Fox, D. W. et al. 2002a, GCN Circ. 1564

\bibitem[Fox et al.(2002b)]{fox02b}
Fox, D. W. et al. 2002b, GCN Circ. 1569

\bibitem[Frail et al.(2001)]{frail01} 
Frail, D.~A.~et al.\ 2001, \apjl, 562, L55 

\bibitem[Garc\'ia-Segura, Mac Low, \& Langer(1996)]{gml96}
Garc\'ia-Segura, G., Mac Low, M.-M., \& Langer, N. 1996, A\&A, 305, 229

\bibitem[Garc\'ia-Segura, Langer, \& Mac Low(1996)]{glm96}
Garc\'ia-Segura, G., Langer, N., \& Mac Low, M.-M. 1996, A\&A, 316, 133

\bibitem[Garcia et al.(1998)]{garcia98} 
Garcia, M.~et al.\ 1998, \apjl, 500, L105 

\bibitem[Garnavich \& Quinn(2002)]{gar02}
Garnavich, P. \& Quinn, J. 2002, GCN Circ. 1661

\bibitem[Garnavich, Loeb, \& Stanek(2000)]{gls00}
Garnavich, P. L., Loeb, A., \& Stanek, K. Z. 2000, ApJ, 544, L11

\bibitem[Gerardy \& Fesen(2001)]{ger01}
Gerardy, C. L. \& Fesen, R. A. 2001, \aj, 121, 2781

\bibitem[Halpern et al.(2002a)]{hal02}
Halpern, J. P. et al. 2002a, GCN Circ. 1578

\bibitem[Halpern et al.(2002b)]{hal02b}
Halpern, J. P. et al. 2002b, GCN Circ. 1593

\bibitem[Hill et al.(1998)]{hill98}
Hill, G.J., Nicklas, H., MacQueen, P.J., Tejada de V., C., Cobos D., F.J.,
\& Mitsch, W. 1998, in Optical Instrumentation, S. D'Odorico, Ed., Proc. 
SPIE, 3355, 375

\bibitem[Hill et al.(2003)]{hill03}
Hill, G.J., Wolf, M.J., Tufts, J.R., \& Smith, E.C. 2003, Proc. SPIE, 
in press

\bibitem[H\"oflich, Khokhlov, \& M\"uller(1992)]{hkm92}
H\"oflich P., Khokhlov A., \& M\"uller M. 1992, A\&A 259, 549

\bibitem[Holland et al.(2002a)]{holland02}
Holland, S. T. et al. 2002a, \aj, 124, 639

\bibitem[Holland et al.(2002b)]{hol02b}
Holland, S. T. et al. 2002b, GCN Circ. 1578

\bibitem[Holland et al.(2002c)]{hol02c}
Holland, S. T. et al. 2002c, GCN Circ. 1597

\bibitem[Klotz et al.(2002a)]{klo02a}
Klotz, A. et al. 2002a, GCN Circ. 1614

\bibitem[Klotz et al.(2002b)]{klo02b}
Klotz, A. et al. 2002b, GCN Circ. 1615

\bibitem[Kudritzki \& Puls(2000)]{kp00}
Kudritzki, R.-P. \& Puls, J. 2000, ARA\&A, 38, 613

\bibitem[Lamb et al.(2002)]{lamb02}
Lamb, D. Q. et al. 2002, GCN Circ. 1600

\bibitem[Landolt(1992)]{landolt92}
Landolt, A. U. 1992, \aj, 104, 340

\bibitem[Lazzati \& Perna(2002)]{laz02}
Lazzati, D. \& Perna, R. 2002, MNRAS submitted, astro-ph/0212105

\bibitem[Lazzati et al.(2002)]{lazzati02}
Lazzati, D., Rossi, E., Covino, S., Ghisellini, G., \& Malesani, D. 2002, 
A\&A, in press, astro-ph/0210333

\bibitem[Lindsay et al.(2002a)]{lin02a}
Lindsay, K., Hartmann, D. H., \& Tassinari, J. 2002a, GCN Circ. 1628

\bibitem[Lindsay et al.(2002b)]{lin02b}
Lindsay, K., Hartmann, D. H., Davis, K., \& Leising, M. 2002, 
GCN Circ. 1638

\bibitem[Lotz(1967)]{lotz67}
Lotz, W. 1967, ApJSupp, 14, 207

\bibitem[Malesani et al.(2002a)]{malesani02}
Malesani, D. et al. 2002a, GCN Circ. 1607

\bibitem[Malesani et al.(2002b)]{mal02b}
Malesani, D. et al. 2002b, GCN Circ. 1645

\bibitem[Masetti et al.(2002)]{mas02}
Masetti, N. et al. 2002, GCN Circ. 1603

\bibitem[Matheson et al.(2002)]{matheson02}
Matheson, T. et al. 2002, submitted to \apjl, astro-ph/0210403

\bibitem[Matsumoto et al.(2002)]{matsumoto02}
Matsumoto, K. et al. 2002, GCN Circ. 1594

\bibitem[Mirabal et al.(2002a)]{mirabal02a}
Mirabal, N. et al. 2002a, \apj, 578, 818

\bibitem[Mirabal et al.(2002b)]{mirabal02b}Mirabal, N., Halpern,
J.~P., Chornock, R., \& Filippenko, A.~V. 2002b, GCN Circ. 1618

\bibitem[Mirabal et al.(2002c)]{mir02c}
Mirabal, N. et al. 2002c, GCN Circ. 1602

\bibitem[M{\o}ller et al.(2002)]{moller02}
M{\o}ller, P. et al. 2002, A\&A, in press

\bibitem[Nakar, Piran, \& Granot(2002)]{npg02}
Nakar, E., Piran, T., \& Granot, J. 2002, New Astronomy, submitted, 
astro-ph/0210631

\bibitem[Oksanen et al.(2002a)]{oks02a}
Oksanen, A., et al. 2002a, GCN Circ. 1570

\bibitem[Oksanen et al.(2002b)]{oks02b}
Oksanen, A. et al. 2002b, GCN Circ. 1591

\bibitem[Panaitescu \& Kumar(2001)]{pk01} 
Panaitescu, A. \& Kumar, P. 2001, \apj, 554, 667

\bibitem[Panaitescu \& Kumar(2002)]{pk02} 
Panaitescu, A. \& Kumar, P. 2002, \apj, 571, 779

\bibitem[Panaitescu, M\'esz\'aros, \& Rees(1998)]{pmr98} 
Panaitescu, A.,  M\'esz\'aros, P., \& Rees, M.J. 1998, \apj, 503, 314

\bibitem[Piro et al.(1999)]{piro99}
Piro, L. et al. 1999, \apj, 514, L73

\bibitem[Piro et al.(2000)]{piro00}
Piro, L. et al. 2000, Science, 290, 955

\bibitem[Pooley(2002)]{pooley02}
Pooley, G. 2002, GCN 1588

\bibitem[Reichart(1999)]{reichart99}
Reichart, D. E. 1999,\apj, 521, L111

\bibitem[Reichart(2001)]{reichart01}
Reichart, D. E. 2001,\apj, 554, 643

\bibitem[Sahu et al.(2002)]{sahu02}
Sahu, D. K. et al. 2002, GCN Circ. 1587

\bibitem[Sako \& Harrison(2002)]{sh02}
Sako, M. \& Harrison, F. 2002, GCN 1624

\bibitem[Salamanca et al.(2002)]{salamanca02}Salamanca, I., Rol., E.,
Wijers, R., Ellison, S., Kaper, L., \& Tanvir, N. 2002, GCN Circ. 1611

\bibitem[Sargent, Steidel, \& Boksenberg(1988)]{ssb88}
Sargent, W. L. W., Steidel, C. C., \& Boksenberg, A. 1988, ApJSupp, 
68, 539

\bibitem[Savaglio et al.(2002)]{savaglio02}
Savaglio, S. et al. 2002, GCN Circ. 1633

\bibitem[Schaefer(2002)]{schaefer02}
Schaefer, B. E. 2002, ApJLett, 583, in press (astro-ph/0212454)

\bibitem[Schaefer, Deng, \& Band(2001)]{sdb01}
Schaefer, B. E., Deng, M., \& Band, D. L. 2001, \apjl, 563, L123

\bibitem[Shirasaki et al.(2002)]{shirasaki02}
Shiraski, Y. et al. 2002, GCN Circ. 1565

\bibitem[Stanek et al.(2002)]{sta02}
Stanek, K. Z., Bersier, D., Winn, J., \& Garnavich, P. 2002, GCN 
Circ. 1598

\bibitem[Stefanon et al.(2002)]{ste02}
Stefanon, M. et al. 2002, GCN Circ. 1623

\bibitem[Uemura et al.(2002)]{uem02}
Uemura, M. et al. 2002, GCN Circ. 1566

\bibitem[Veigele(1973)]{veigele73}
Veigle W.J. 1973, Atomic Data Tables 5, 51

\bibitem[Vietri \& Stella(1998)]{vs98}
Vietri, M. \& Stella, L. 1998, \apj, 507, L45

\bibitem[Vreeswijk et al.(2001)]{vreeswijk01}
Vreeswijk, P. M. et al. 2001, \apj, 546, 672

\bibitem[Wang \& Loeb(2000)]{wangloeb00}
Wang, X. \& Loeb, A. 2000, \apj, 535, 788

\bibitem[Weidinger et al.(2002)]{wei02}
Weidinger, M. et al. 2002, GCN Circ. 1573

\bibitem[Wheeler, Yi, H{\" o}flich, \& Wang(2000)]{wheel00}
Wheeler, J.~C., Yi, I., H{\" o}flich, P., \& Wang, L.\ 2000, \apj, 537,
810 

\bibitem[Williams et al.(2002)]{wil02}
Williams, G., Lindsay, K., \& Milne, P. 2002, GCN Circ. 1652

\bibitem[Winn et al.(2002)]{win02}
Winn, J. et al. 2002, GCN Circ. 1576

\bibitem[Woosley \& Heger(2001)]{woo01} 
Woosley, S. E. \& Heger, A. 2001, AAS, 198, 3801

\bibitem[Woosley, Langer, \& Weaver(1993)]{wlw93}
Woosley, S. E., Langer, N., \& Weaver, T. A. 1993, \apj, 411, 823

\bibitem[Zharikov et al.(2002)]{zha02}
Zharikov, S., Vazquez, R., Benitez, G., \& del Rio, S. 2002, GCN 
Circ. 1577

\end{thebibliography}
\end{document}